\documentclass[%
 reprint,
 amsmath,amssymb,
prb,
floatfix,
]{revtex4-1}

\usepackage{graphicx}% Include figure files
\usepackage{dcolumn}% Align table columns on decimal point
\usepackage{bm}% bold math
\begin{document}

\preprint{APS/123-QED}

\title{Two\textendash photon lasing by a superconducting qubit}

\begin{abstract}
We study the response of a  magnetic-field-driven superconducting qubit strongly coupled to a superconducting
coplanar waveguide resonator. We observed a strong amplification/damping of a probing signal at different resonance points corresponding to a one and two-photon emission/absorption. The sign of the detuning between the qubit frequency and the probe determines whether amplification or damping is observed. The larger blue detuned driving
leads to two-photon lasing while the larger red detuning cools the resonator.
Our experimental results are in good agreement with the theoretical model of qubit lasing and cooling at the Rabi frequency.
\end{abstract}

\date{\today}
\author{P. Neilinger}
\author{M. Reh\'{a}k}
\author{M. Grajcar}
\affiliation{Department of Experimental Physics, Comenius University, SK-84248 Bratislava, Slovakia}
\affiliation{Institute of Physics, Slovak Academy of Sciences, 845 11 Bratislava, Slovakia}
\author{G. Oelsner}
\author{U. H\"{u}bner}
\affiliation{Leibniz Institute of Photonic Technology, P.O. Box 100239, D-07702 Jena, Germany}
\author{E. Il'ichev}
\affiliation{Leibniz Institute of Photonic Technology, P.O. Box 100239, D-07702 Jena, Germany}
\affiliation{Novosibirsk State Technical University, 20 K. Marx Ave., 630092 Novosibirsk, Russia}
\pacs{\ldots}
\maketitle

Motivated by the first experiment demonstrating the energy exchange between a strongly driven superconducting qubit and a resonator at the Rabi frequency $\Omega_R$\cite{Ilichev03}, Hauss et al.\cite{Hauss2008} elaborated a theoretical model 
to quantify this phenomenon. Their model predicts large resonant effects for the one- and two-photon resonance conditions $\Omega_R = \omega_r-g_3\bar{n}$  and $\Omega_R = 2\omega_r-g_3\bar{n}$, where  $\omega_r$ is the fundamental frequency of the resonator, $g_3$ is the effective coupling energy, and $\bar{n}$ is the average number of photons in the resonator at frequency $\omega_r$. Depending on the detuning between the driving frequency $\omega_d$ and the qubit eigenfrequency $\omega_q$, either a lasing behavior (blue detuning $\omega_d$ - $\omega_q > 0$) of the oscillator can be realized or the qubit can cool the oscillator (red detuning $\omega_d$ - $\omega_q < 0$). According to the theory, one-photon lasing/cooling effects vanish at the symmetry (degeneracy) point of the qubit. However, the two-photon processes persist at the symmetry point where the qubit-oscillator coupling is quadratic and decoherence
effects are minimized. There, the system realizes a “single-atom-two-photon laser”.
Note a similar two-photon lasing by a quantum dot in a microcavity, which was  investigated theoretically in Ref.~\onlinecite{Valle10}.

Experimentally, a single qubit one-photon lasing was demonstrated by the NEC group \cite{Astafiev2007}. Here, a single charge qubit was used and a population inversion was provided by single-electron tunneling. Later on, the amplification/deamplification of a transmitted signal trough a coplanar waveguide resonator was achieved by a strongly driven single flux qubit.\cite{Oelsner13} However, the two-photon lasing has not been experimentally demonstrated yet.
In this paper, we demonstrate the two-photon lasing, as well as  considerable enhancement of one-photon lasing of a superconducting qubit by one order of magnitude in comparison with Ref.~\onlinecite{Oelsner13}. This enhancement was achieved by a much stronger coupling of the superconducting qubit to the resonator.

\begin{figure}[h]
\centering \includegraphics[width=7.5cm]{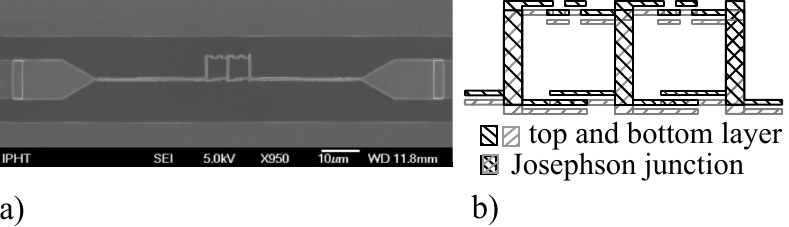}
\caption{a) Scanning electron microscope image of the qubits incorporated into the coplanar waveguide resonator. b) Detailed scheme of the qubits. The qubits share a Josephson junction with each other as well as with the resonator.}
\label{fig:2QubitScheme}
\end{figure}

The lasing effect was investigated by making use of a standard arrangement: a superconducting qubit placed in the middle of a niobium $\lambda/2$ coplanar waveguide resonator. The latter was fabricated by conventional sputtering and dry etching of a 150-nm-thick niobium film. The patterning uses an electron beam lithography and a CF$_4$ ion etching process. The aluminum qubits were fabricated by the shadow evaporation technique. The coupling between the qubit and the resonator was implemented by a shared Josephson junction (Fig.~\ref{fig:2QubitScheme}). The dimensions of the qubit's Josephson junctions are
$0.2\times 0.3~\mu$m$^2$, $0.2\times 0.2~\mu$m$^2$ and $0.2\times 0.3~\mu$m$^2$, the critical current density is about 200 A/cm$^{2}$,
and the area of the qubit loop is $5\times 4.5~\mu$m$^2$.
The resonance frequency and the quality factor of the resonator's fundamental mode taken for a weak probing (-141~dBm) are $\omega_r = 2\pi \times 2.482$~GHz, $Q_0=18\;000$. The same parameters of the third harmonics taken at the same power are $\omega_{r3}=2\pi\times 7.446$~GHz, $Q_3=3750$. These values were determined from the transmission spectra of the coplanar waveguide resonator.

%and were designed to be identical sharing a coupling Josephson junction with dimensions $1\times5~\mu$m$^2$. This junction provides coupling between qubits \cite{Grajcar05,Grajcar06}.

%We designed such unit cell of two qubit with the aim to form a metamaterial with magnetically tunable parameters by simple extension of the unit cell into an one-dimensional array incorporated into a coplanar waveguide.\cite{Rehak14} In such metamaterial a sufficient  amplification can be achieved even for traveling microwaves waves providing large bandwidth of the microwave amplifier.In this work we characterize the properties of the  unit cell incorporated into the coplanar waveguide resonator from the resonator transmission measurements which enable to determine all its important parameters.

In practice, we measured a two-qubit sample which represents a unit cell of a one-dimensional array of ferromagnetically coupled qubits exhibiting a large Kerr nonlinearity.\cite{Rehak14} However, by applying a certain energy bias, one qubit can be set to a localized state, while the second is in the vicinity of its degeneracy point. This way, we can measure the qubits separately to reconstruct their parameters\cite{Ilichev04}, and the dynamics of the system is defined by a single qubit only.
Therefore, to describe our findings, we will use the one-qubit model elaborated in Ref.~\onlinecite{Hauss2008}, in which the corresponding Hamiltonian reads in the flux basis of the qubit:
\begin{eqnarray}
	H&=&-\frac{1}{2}\epsilon\sigma_z-\frac{1}{2}\Delta\sigma_x-\hbar\Omega_{R0}\cos(\omega_dt)\sigma_z\\
	&+& \hbar\omega_ra^\dagger a + g\sigma_z(a^\dagger+a)\nonumber
	\label{eq:H}
\end{eqnarray}
where $\Delta$ is the energy level separation of the two level system at zero energy bias $\epsilon=0$, $\Omega_{R0}$ is the driving amplitude of the applied microwave magnetic flux with frequency $\omega_d$, and $g$ is the coupling energy between the qubit and the resonator. The coupling energy scales with the ratio of the magnitude of the persistent current in the qubit  $I_q$ and   the critical current of the coupling Josephson junction $I_{c0}$ as
\begin{equation}
g\equiv\hbar\omega_g=\frac{\hbar\omega_r}{2\pi}\frac{I_q}{I_{c0}}\sqrt{\frac{1}{G_0Z_r}}
\end{equation}
where $Z_r=50\Omega$ is the wave impedance of the coplanar waveguide resonator and $G_0\equiv 2e^2/h$ is the quantum conductance.

This Hamiltonian can be transformed  by the
Schrieffer-Wolff transformation $U=\exp(iS)$ with the generator
$S=(g/\hbar\omega_q)\cos\eta(a+a^\dagger)\sigma_y$
and a rotating wave approximation $U_R=\exp(-i\omega_d\sigma_zt/2)$ to the Hamiltonian\cite{Hauss2008}
\begin{eqnarray}
	\tilde{H}&=&\hbar\omega_ra^\dagger a +\frac{1}{2}\hbar\Omega_R\sigma_z \nonumber \\
	&+& g\sin\eta[\sin\beta\sigma_z-\cos\beta\sigma_x](a+a^\dagger) \nonumber \\
	&-& \frac{g^2}{\hbar\omega_q}\cos^2\eta[\sin\beta\sigma_z-\cos\beta\sigma_x](a+a^\dagger)^2 \nonumber
\label{eqH_}
\end{eqnarray}
Here, $\hbar\omega_q=\sqrt{\epsilon^2+\Delta^2}$,
$\Omega_R=\sqrt{\Omega_{R0}^2\cos^2\eta+\delta\omega^2}$, $\tan\eta=\epsilon/\Delta$, $\tan\beta=\delta\omega/(\Omega_{R0}\cos\eta)$ and $\delta\omega=\omega_d-\omega_q$. 
The transmission of the resonator
\begin{equation}
t\propto\langle a\rangle
\end{equation}
where
\begin{equation}
\langle a\rangle=tr(\tilde{\rho} a)
\end{equation}
was calculated numerically by the quantum toolbox Qutip,\cite{Qutip} solving the Liouville equation for the density matrix of the system in the rotating frame
\begin{equation}
\dot{\tilde{\rho}}=-\frac{i}{\hbar}\left[\tilde{H},\tilde{\rho}\right]+\tilde{L}_q\tilde{\rho}+\tilde{L}_r\tilde{\rho}
\end{equation}
where $\tilde{L}_q$ and $\tilde{L}_r$ are Lindblad superoperators
\begin{eqnarray}
\tilde{L}_q \tilde{\rho}&=&
\frac{\tilde{\Gamma}_\downarrow}{2}(2\sigma_-\tilde{\rho}\sigma_+-\tilde{\rho}\sigma_+\sigma_--\sigma_+\sigma_-\tilde{\rho}) \nonumber \\
&+&\frac{\tilde{\Gamma}_\uparrow}{2}(2\sigma_+\tilde{\rho}\sigma_--\tilde{\rho}\sigma_-\sigma_+-\sigma_-\sigma_+\tilde{\rho}) \nonumber \\
&+&\frac{\tilde{\Gamma}_\varphi}{2}(\sigma_z\tilde{\rho}\sigma_z-\tilde{\rho}),
\end{eqnarray}
\begin{eqnarray}
\tilde{L}_r \tilde{\rho} &=&
\frac{\kappa}{2}(N_{th}+1)(2a\tilde{\rho} a^\dagger-\tilde{\rho} a^\dagger a-a^\dagger a\tilde{\rho}) \nonumber \\
&+&\frac{\kappa}{2}N_{th}(2a^\dagger \tilde{\rho} a-a a^\dagger \tilde{\rho}-\tilde{\rho} a a^\dagger ).
\end{eqnarray}
Here $N_{th}= 1/\,[exp(\hbar \omega_T/k_B T)-1]$  is the thermal distribution function of photons in the resonator, $\kappa$ is resonator loss rate, $\tilde{\Gamma}_{\downarrow , \uparrow}$ and $\tilde{\Gamma}_\varphi$ are the relaxation, excitation and dephasing rates in the rotating frame derived in Ref.~\onlinecite{Hauss2008}

\begin{eqnarray}
\tilde{\Gamma}_{\uparrow ,\downarrow}&=&
\frac{\Gamma_0}{4}\cos^2\eta(1\pm\sin\beta)^2+
\frac{\Gamma_\varphi}{2}\sin^2\eta\cos^2\beta \nonumber \\
\tilde{\Gamma}_\varphi &=& \frac{\Gamma_0}{2}\cos^2\eta\cos^2\beta+
\Gamma_\varphi\sin^2\eta\sin^2\beta
\end{eqnarray}

\begin{figure}[h!]
\centering
\includegraphics[width=8.0cm]{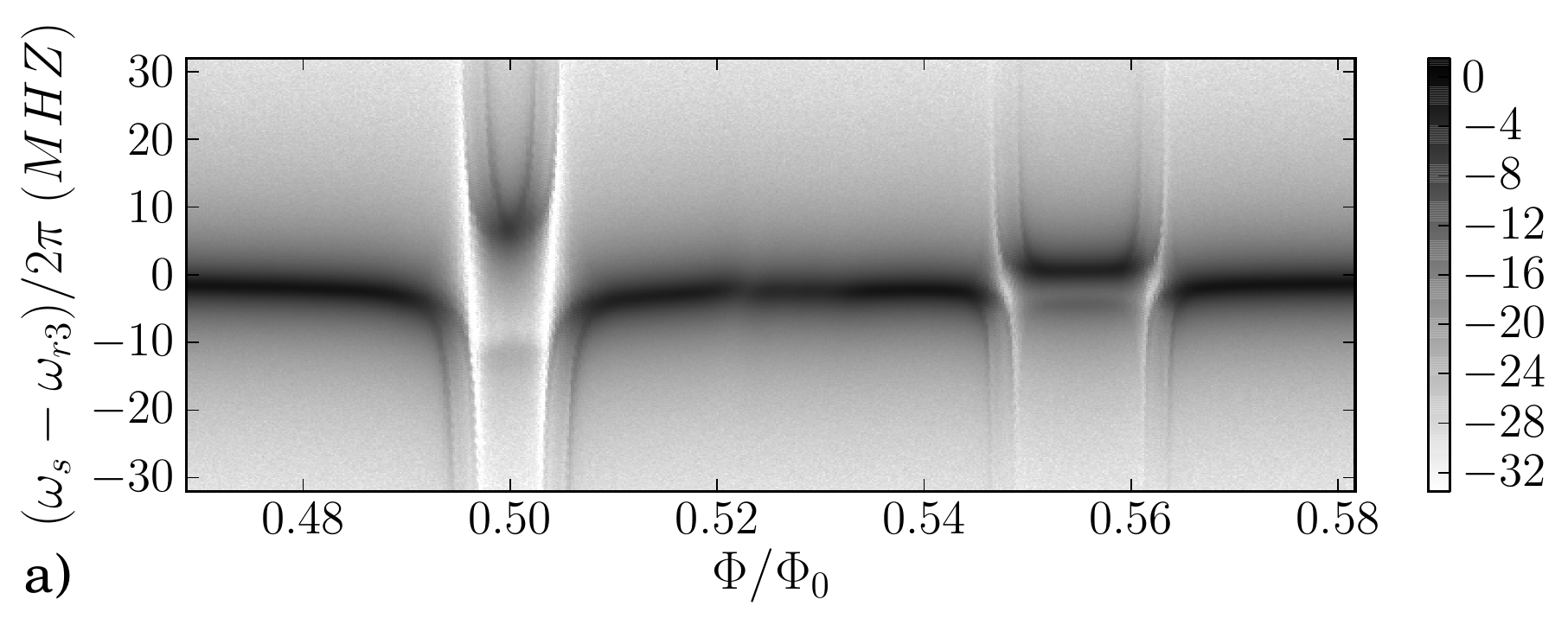}
\includegraphics[width=8.0cm]{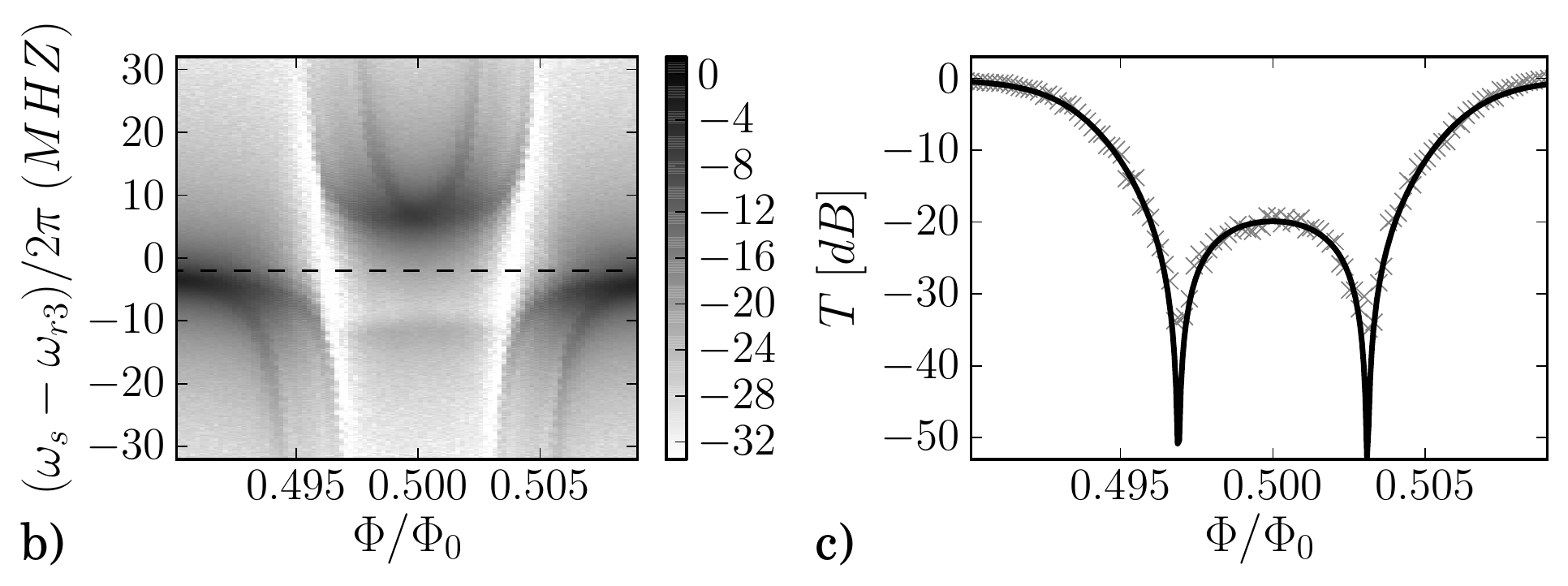}
\caption{ (a) Resonator transmission as a function of the magnetic flux and the detuning of the resonator from the third harmonic resonance frequency $\omega_{r3}/2\pi$. The anticrossings of qubit A and B are separated by a magnetic flux of about $55\times 10^{-3}\Phi_0$. (b) Close view of the transmission in the vicinity of the left qubit's (A) degeneracy point and (c) the cut of the transmission map along the dashed line in (b).  The crosses are experimental data and the solid line is a theoretical curve calculated from Eq.~\ref{eq:twd}.}
\label{fig:75fity}
\end{figure}

The qubit parameters used for the numerical calculations were determined independently from the transmission of the resonator $t$ coupled to the undriven qubit. For a weak microwave signal with frequency $\omega_s$, the transmission can be expressed in simple form \cite{Omelyanchouk10}

\begin{equation}
t=-i\dfrac{\kappa_{ext}}{2}\dfrac{\delta\omega_q+i\gamma}
{\omega_g^2\cos^2\eta-\left(\delta\omega_r+i\kappa/2\right)
\left(\delta\omega_q+i\gamma\right)}
\label{eq:twd}
\end{equation}

where $\delta\omega_q=\omega_q-\omega_s$, $\delta\omega_r=\omega_r-\omega_s$, $\kappa_{ext}$ is the external loss rate of the resonator and $\gamma$ is the qubit decoherence rate. The experimental data was fitted by Eq.\ref{eq:twd} (see Fig.~\ref{fig:75fity}) and the qubit parameters
obtained from the fitting procedure are given in the table~\ref{tab:qubit_par}.

\begin{table}[h!]
\centering
    \begin{tabular}{ccccc}
    \toprule
    Qubit & I$_p$   & $\Delta$/2$\pi$ & g/2$\pi$   & $\gamma$/2$\pi$ \\
    $\ $      & (nA)&(GHz)     &(MHz)&(MHz)      \\
    A     & 208 & 6.39     & 109 & 15      \\
    B     & 138 & 5.28     & 77  & 20      \\
    \end{tabular}
\caption{Qubit parameters determined from the fitting procedure.}
\label{tab:qubit_par}
\end{table}

\begin{figure}[h!]
\centering
\includegraphics[width=7.0cm]{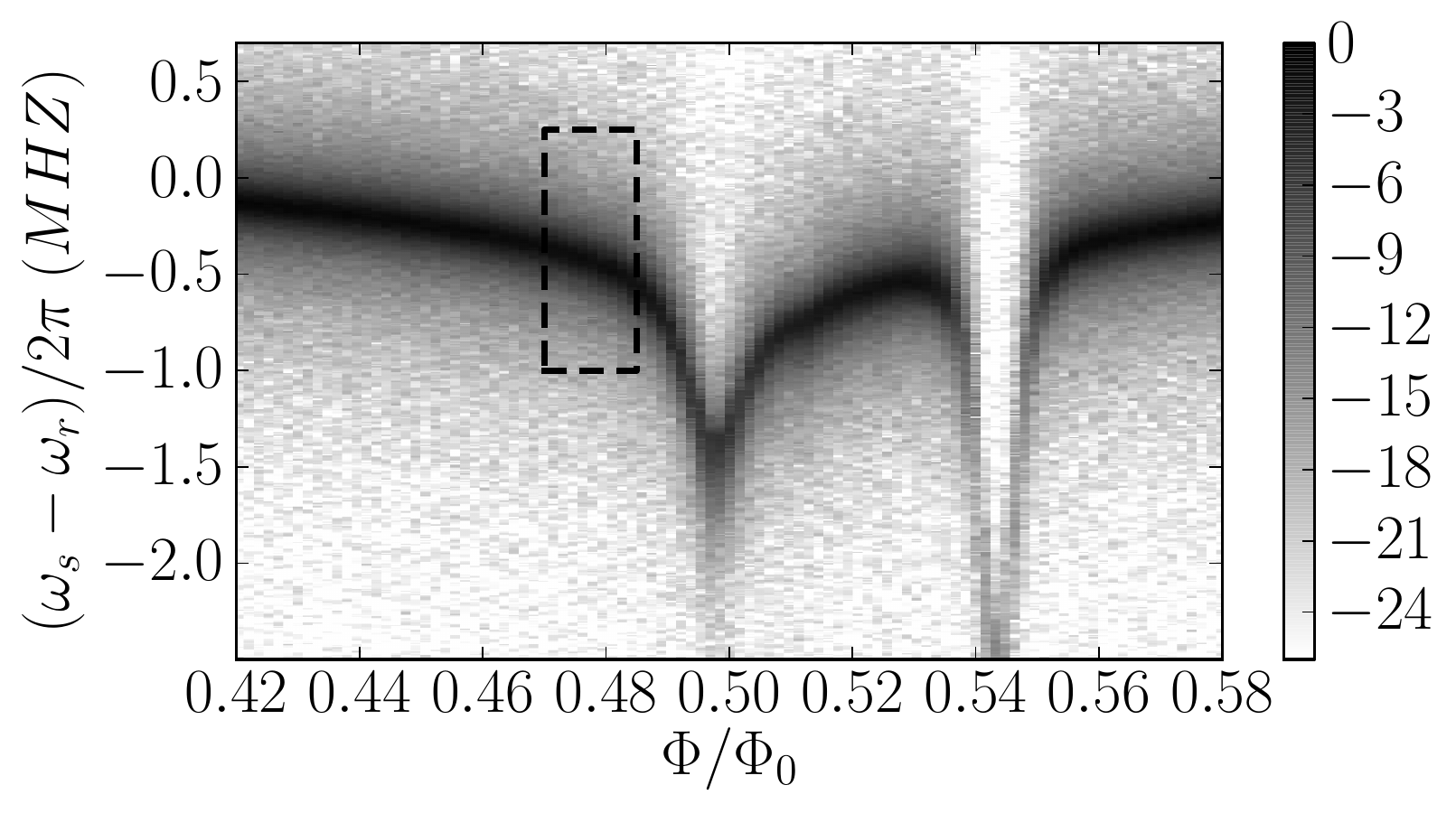}
\caption{a) Resonator transmission in dB units as a function of the magnetic flux and the detuning of the resonator from the resonance frequency 2.482~GHz.}
\label{fig:ABDisp}
\end{figure}

\begin{figure}[h!]
\centering
\includegraphics[width=7.0cm]{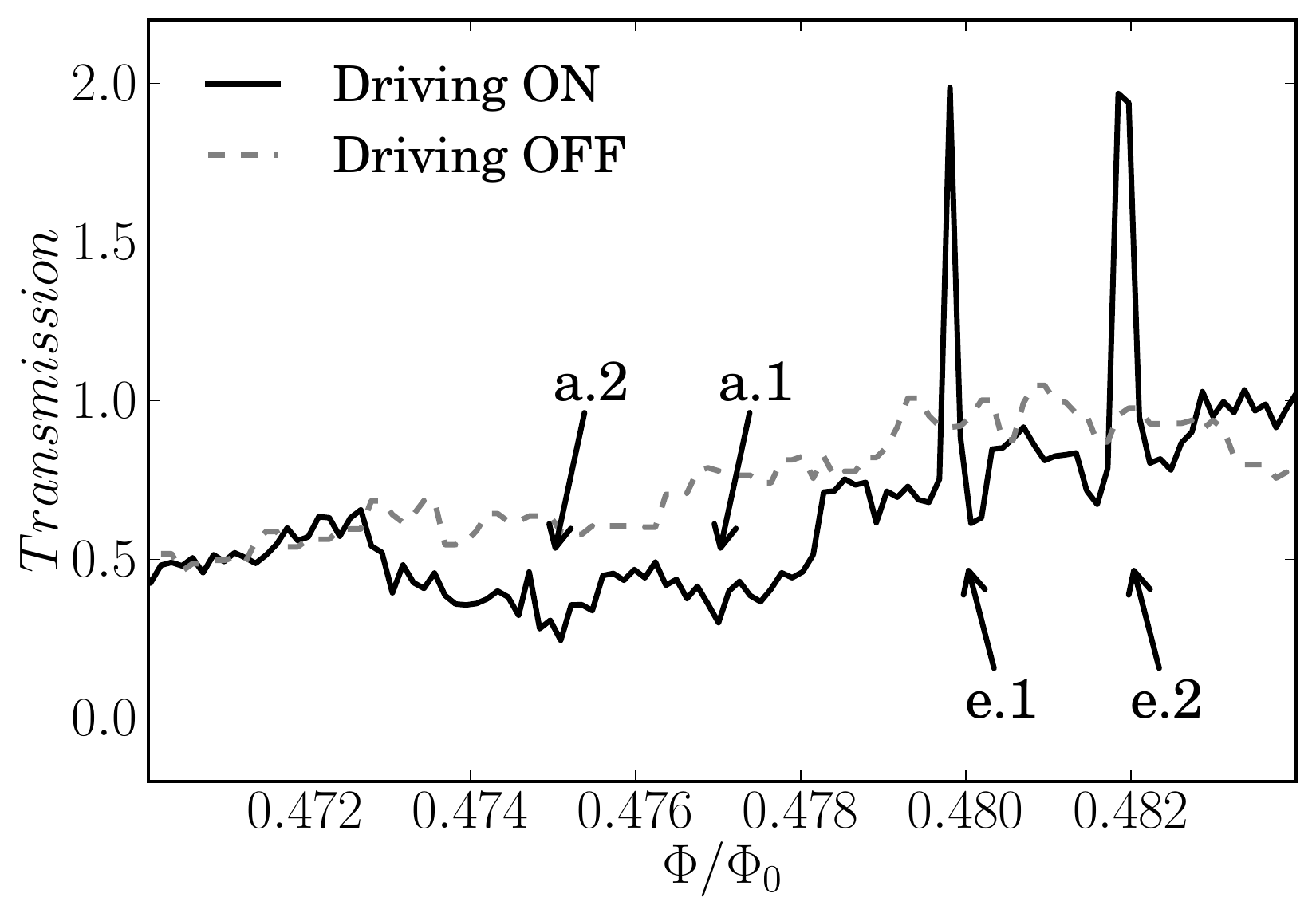}
\includegraphics[width=7.0cm]{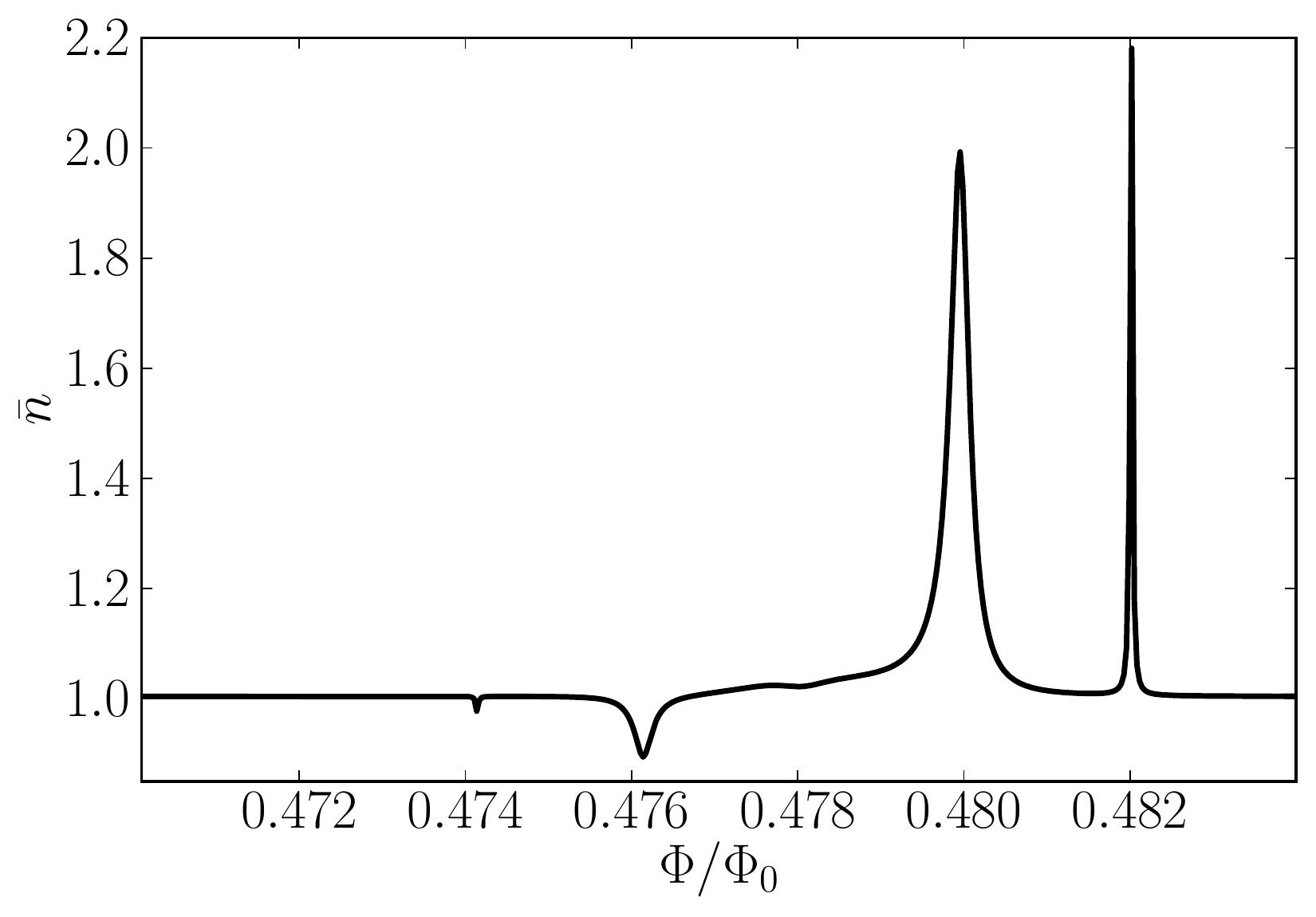}
\caption{ Transmission of the resonator at fixed frequency $\omega_s/2\pi$ for a driving signal with frequency $\omega_d/2\pi=9\omega_r/2\pi=22.338$~GHz and power  $p_d=-102$~dBm for the driving switched off and on (a). Regions $e.1$, $e.2$ and  $a.1$, $a.2$ exhibit amplification and attenuation of the signal, respectively. Panel (b) shows the simulated average photon number in the resonator obtained for the qubit parameters.}
\label{fig:LasingE}
\end{figure}

We have investigated the stimulated emission effect observed when strongly driving the system at a frequency $\omega_d/2\pi=9\omega_r/2\pi=22.338$~GHz for qubit A.
The resonator transmission was measured by a network analyzer at resonance $\omega_r$
for magnetic fluxes marked by the black rectangular area in Fig.~\ref{fig:ABDisp} and is shown in Fig.~\ref{fig:LasingE}a.
At a driving power $P_d=-103$~dBm, two emission peaks ($e.1$, $e.2$) accompanied by two attenuation dips ($a.1$, $a.2$) appear in the transmission spectra.
The increase of the transmission is accompanied by a narrowing of the resonance curve. The one-photon  ($a.1$, $e.1$) and two photon ($a.2$, $e.2$) processes are enhanced at resonance with the Rabi frequency of the qubit $\Omega_R = \omega_r-g_3\bar{n}$ and $\Omega_R = 2\omega_r-g_3\bar{n}$, respectively. These results are in good agreement with the theoretical
model\cite{Hauss2008} described above for parameters given in Table~\ref{tab:qubit_par}.
By a strong coupling of the qubit to the resonator, we have achieved a considerable enhancement of the lasing, nearly one order of magnitude, in comparison with the results presented in Ref.~\onlinecite{Oelsner13}.
Further improvement is possible by increasing the relaxation rate of the qubit, for instance, by placing a gold resistor close to the qubit loop.

To conclude, we have experimentally demonstrated single-qubit one-photon and two-photon lasing. The experimental
results are in good agreement with the theoretical model developed by Hauss et al.\cite{Hauss2008}
The considerable enhancement of lasing effect was achieved by stronger coupling of the superconducting qubit to the resonator, and theoretical calculations show that it can be enhanced further by increasing the relaxation rate of the qubit.
Such improvement could enable to observe even higher-order processes analysed in Ref.~\onlinecite{Shevchenko2014}.

\begin{figure}[h!]
\centering
\includegraphics[width=7.0cm]{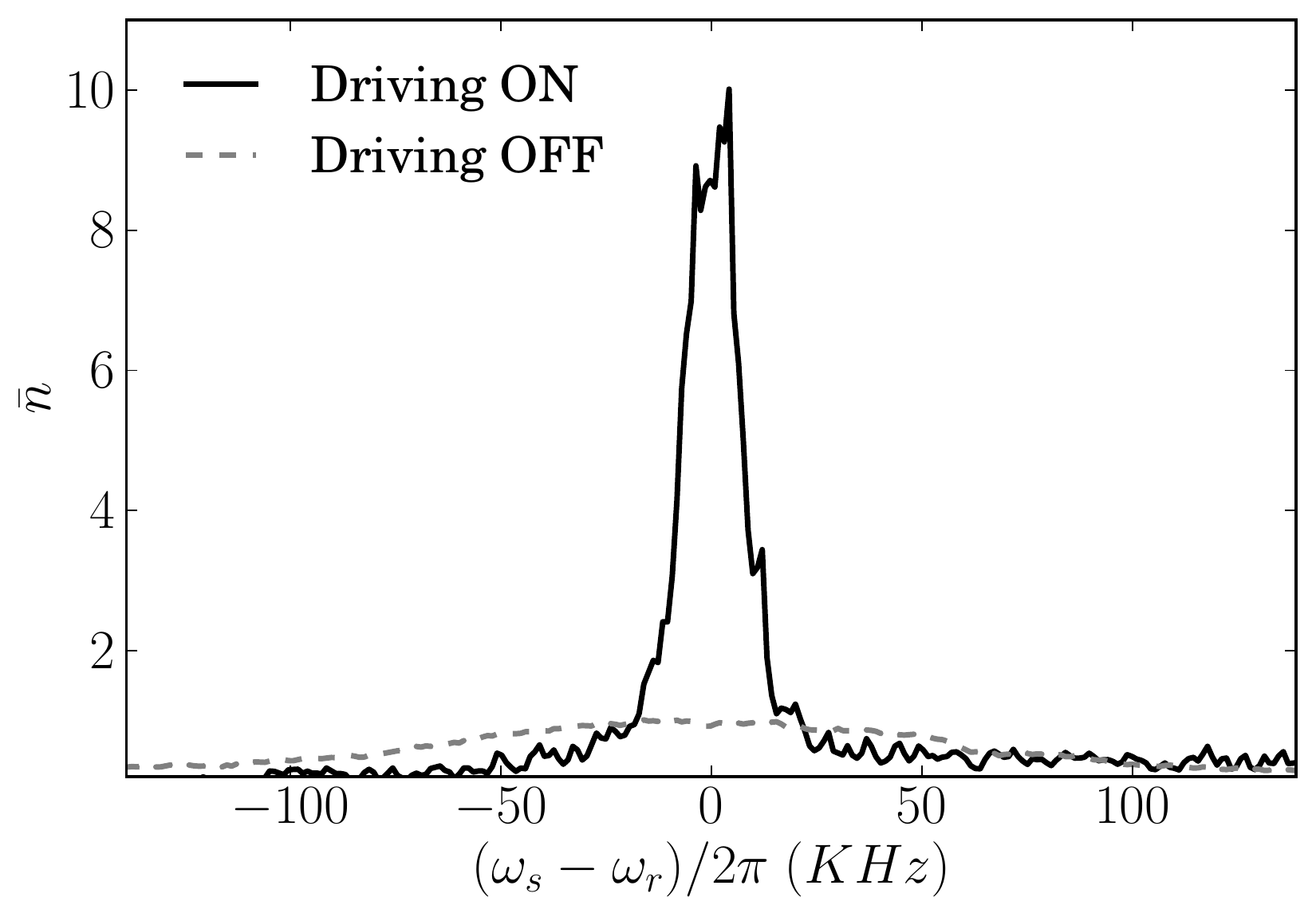}
\caption{Resonance curve of the resonator at driving switched off (solid line) and on (dashed line) with frequency $\omega_d/2\pi=9\omega_r/2\pi=22.338$~GHz and power $p_d=-101$~dBm in region $e.1$. The amplitude increases by a factor of $\sim 9$ and bandwidth is reduced by factor of 10.}
\label{fig:LasingED}
\end{figure}

The research leading to these results has received funding from the European Community’s
Seventh Framework Programme (FP7/2007-2013) under Grant No. 270843
(iQIT) and APVV-DO7RP003211.
This work was also supported by the Slovak Research and Development Agency
under the contract APVV-0515-10 and APVV-0808-12(former projects No. VVCE-0058-07, APVV-0432-07)
and LPP-0159-09. The authors gratefully acknowledge the financial support of the EU through the ERDF OP R$\&$D, Project CE QUTE $\&$ metaQUTE, ITMS: 24240120032 and CE SAS QUTE. EI acknowledges partial support of
Russian Ministry of Education and Science, in the framework
of state assignment 8.337.2014/K.

\bibliographystyle{apsrev4-1}
\bibliography{references}% Produces the bibliography via BibTeX
\end{document}